\documentclass[iop]{emulateapj}

\begin{document}

\title{The Discovery of Solar-like Activity Cycles Beyond the End of the Main Sequence?}

\author{Matthew Route\altaffilmark{1}}

\altaffiltext{1}{Research Computing, Information Technology at Purdue, Purdue University, 155 S. Grant St., West Lafayette, IN 47907, mroute@purdue.edu}

\slugcomment{Accepted for publication in ApJL: 2016 September 22}

\begin{abstract}

The long-term magnetic behavior of objects near the cooler end of the stellar main sequence is poorly understood.  Most theoretical work on the generation of magnetism in these ultracool dwarfs (spectral type $\geq$M7 stars and brown dwarfs) suggests that their magnetic fields should not change in strength and direction.  Using polarized radio emission measurements of their magnetic field orientations, I demonstrate that these cool, low-mass, fully-convective objects appear to undergo magnetic polarity reversals analogous to those that occur on the Sun.  This powerful new technique potentially indicates that the patterns of magnetic activity displayed by the Sun continue to exist, despite the fully convective interiors of these objects, in contravention of several leading theories of the generation of magnetic fields by internal dynamos. 

\end{abstract}

\section{Introduction}

The Sun undergoes cycles in magnetic activity, as manifested through the $\sim$11-year sunspot cycle, the associated periodic alterations in solar flux output at X-ray, ultraviolet, optical, and radio wavelengths, and reversals in magnetic field polarity of both sunspots and the large-scale dipolar field \citep{hat10}.  One full magnetic cycle consists of two sunspot cycles and a reversal in magnetic field direction in each hemisphere, followed by the return to their original states.  Solar active regions contain sunspots, which can occur in isolation, in pairs, and in groups \citep{sol03}.  When sunspots appear in pairs, each sunspot in the pair is usually of opposite magnetic polarity, with the same polarity leading the sunspot pairs in the northern hemisphere, and the opposite polarity leading pairs in the southern hemisphere.  The leading sunspot's polarity matches the polarity of the large-scale dipolar magnetic field in that hemisphere, though the phase of the polarity lags that of the active regions by half of a sunspot cycle.  Additionally, ``Joy's Law'' describes how the trailing sunspot in a sunspot pair is at a higher latitude, and that the higher the latitude, the greater the inclination of the sunspot pair axis to the equator \citep{cha14}.  As the solar activity cycle progresses, sunspots and active regions emerge near mid-latitudes and progressively approach the equator, while their overall numbers increase, as exhibited by the ``Butterfly'' diagram \citep{hat10}.  The large-scale, dipolar solar magnetic field features a stronger poloidal (but weaker toroidal) field during solar minimum, and a stronger toroidal (but weaker poloidal) field during solar maximum \citep{hat10,mci14}.

Observations of the majority of F2-M2 main sequence stars indicate that they also display such cycles \citep{bal95,ola09}.  Beyond spectral type $\sim$M3, the interiors of stars become fully convective \citep{cha97,cha00}, which is thought to disrupt the production of large-scale magnetic fields by the $\alpha\Omega$ dynamo mechanism \citep{cha14} present in the Sun and other stars with convective envelopes.  At spectral types $\geq$M7, the usual indicators of magnetic activity, H$\alpha$ and X-rays, decline in these ultracool dwarfs (UCDs) \citep{mcl12} presumably due to their increasingly neutral atmospheres \citep{moh02}, yet radio emission still permits a diagnosis of the magnetic fields of objects as cool as spectral type T6.5 ($\sim$900 K) \citep{rou12}.

Section 2 describes how long-term observations of radio flares from UCDs may be used to trace the evolution of their magnetic field orientations and topologies.  Section 3 discusses the implication of these results on detection statistics, and more importantly, on the evaluation of the dynamo models of their interiors.

\section{Radio Observations of Ultracool Dwarf Magnetism}

During observations of the T6.5 brown dwarf 2MASS J10475385+2124234 (J1047+21) conducted by the Arecibo radio telescope from 2010 January to 2011 February, three radio flares were detected at a center frequency of 4.75 GHz \citep{rou12}.  Two of the flares featured circular polarization fractions $>$70\%, and all were right circularly polarized (RCP) with brightness temperatures $>$ 5 x 10$^{10}$ K.  These radio flares were most likely caused by the electron cyclotron maser instability (ECMI) on account of their high circular polarization fractions and large, non-thermal brightness temperatures that point to a coherent emission mechanism. Such radio bursts are generated by other UCDs, the Solar System planets, and potentially the Sun \citep{hal06,tre06}.  This emission mechanism operates at the local cyclotron frequency ($\nu_{c} = 2.8 \times 10^{6}B$ G), yielding a B$\geq$1.7 kG[auss] magnetic field in the emitting region of J1047+21.

A recent follow-up survey of UCDs by the Arecibo radio telescope detected an additional flare from J1047+21, occurring in 2013 May that is also RCP, and exhibits properties similar to the previously detected flares \citep{rou16b}.  However, each and every flare detected in 2013 December \citep{wilb15} is nearly 100\% left circularly polarized (LCP), reflecting a change in the emitting region magnetic field polarity.  This is unexpected, since repeated radio observations of active regions on nearby active M dwarfs previously indicated unchanging magnetic field orientations, suggesting the presence of stable, global, dipolar fields \citep{whi02}. That the flaring emission should change such that all flares exhibit one polarity during a given time period, then at some later time all exhibit the opposite polarity, could indicate reversals in the kG-strength magnetic fields present in the emitting regions, and are suggestive of the global magnetic reversal (polarity transition) phenomenon.

Inspired by the possibility that J1047+21 may have undergone a global dipolar magnetic field reversal as part of a solar-like activity cycle, I investigated the long-term magnetic behavior of all known radio-flaring UCDs, as summarized in Table 1.  The flaring L1.5 brown dwarf of the binary 2MASS J07464256+2000321 system \citep{ber09,har13} which has been observed over 7 years, appears to have undergone one complete activity cycle, featuring two polarity reversals.  From data collected in 2007 Jan \citep{ant08} and 2008 Feb \citep{ber09}, the radio emission is observed to flare repeatedly, but the helicity of the polarization is always highly LCP.  Yet in observations stretching from 2010 Nov to 2010 Dec, only RCP radio flares were detected, followed by a gradual change to both LCP and RCP flares on 2010 Dec 16 \citep{lyn15}.  Subsequent observations from 2012 Feb to 2012 Dec only detect highly LCP radio emission \citep{rouphd,lyn15}.  The smooth, systematic transition from one type of magnetic field polarity to another may indicate the formation and dissipation of active regions with small covering fractions that migrate into and out of view during the course of an activity cycle.

The radio emission from 2MASS J13142039+1320011 AB supports the hypothesis of a magnetic activity cycle in two distinct ways.  In 2009 March, low-polarized, presumably gyrosynchrotron, emission was detected, but no radio bursts were reported despite the instrument sensitivity and observation duration to record multiple flares \citep{mcl11,wil15}.  This quiescent radio emission slowly varied sinusoidally between $\pm$24\% circular polarization over the rotational period of the object.  Yet between 2012 March and 2013 May, highly LCP flares were reported \citep{wil15,rou16b}.  This alteration between global quiescence and activity suggests that a stellar activity cycle is at work, where the quiescent period represents a solar minimum phase during the activity cycle.

LSR J1835+3259 also exhibits flaring behavior consistent with one magnetic polarity transition (one-half of an activity cycle), including a quiescent period \citep{ber08b}, although the observational data are sparse.  The newly-detected T6-dwarf, WISEPC J112254.73+255021.5, with an apparent rotational period as little as 0.288 hours (and subharmonics thereof) has only been observed to emit LCP emission over a timescale of $\sim$8 months, and thus has not been observed long enough to provide evidence for a global magnetic polarity reversal \citep{rou16a}.  Six radio-flaring sources, DENIS 1048-3956 \citep{bur05}, LP 944-20 \citep{ber01}, BRI 0021-0214 \citep{ber02}, 2MASS 10430758+2225236, 2MASS 01365662+0933473, and 2MASS 12373919+6526148 \citep{kao16} have not been observed over multiple epochs, and as such, it is unknown whether they are periodic radio-emitters.  SDSS 04234858-0414035 is a binary source and it is unclear at this time which component(s) may be responsible for its flaring behavior.

On the other hand, radio observations of 2MASS J0036159+182110 (J0036+18) are inconclusive since no polarity reversal is noted in radio observations found in the literature, yet it also may not be considered to have ``flaring" activity, as the radio emission occurs over a timescale of $\sim$1 hr \citep{ber05} as opposed to the rapid, several-minute long flares exhibited by the other listed objects.  However, alterations in another magnetic activity indicator, H$\alpha$, which on the Sun is a telltale sign of activity regions and plages \citep{rob13}, do support the presence of an activity cycle \citep{rou16b}.  The lack of flaring radio behavior and H$\alpha$ emission could be indicative of a brown dwarf resting in a solar minimum phase, whereas the increase in H$\alpha$ emission could signal the resumption of magnetic activity.

Perhaps the best studied UCD at the boundary between cool late-type M dwarf stars and brown dwarfs is the M9 dwarf TVLM 513-46546 (TVLM 513), which exhibits rapid bursts of radio emission of either polarity as little as minutes apart \citep{hal06,rouphd}.  This behavior is consistent with an inclination angle of $\sim$90$^{\circ}$ to the observer \citep{ber08a}, that permits observation of flaring activity in both hemispheres, or from both poles.  Moreover, the potential measurement of differential rotation in TVLM 513 \citep{wol14}, the first for any UCD, is consistent with the equator-ward movement of flaring active region(s) and resembles the behavior of sunspots on the Sun during the sunspot cycle \citep{hat10}.  The analysis of the systematic shortening in the time interval between radio bursts from TVLM 513 over several years permits the calculation of the time for active regions to slowly migrate during an activity cycle from higher latitudes toward the equator.  The time interval from the longest once-per-revolution flaring period, through systematically shorter periods, to the next longest flaring period, yields the length of the starspot cycle, which is one-half the magnetic activity cycle, and can be computed to be 3-4 years, although the UCD has only been observed for $\sim$14 years with large temporal gaps in the radio measurements that prevent the decisive determination of an activity cycle period.  Thus, among the seven UCDs for which periodic radio-flaring behavior is observed, six show indications of stellar-like activity cycles (Table 1).  These results suggest that the timescale for a UCD magnetic activity cycle is $\sim$2-20 years in duration, with a change in polarity occurring every $\sim$1-10 years.

Several models may account for the periodic reversals in magnetic polarity witnessed, or, in the case of TVLM 513, the observation of radio flaring from both magnetic field orientations minutes apart. These include: the presence of large chromospheric hotspot(s) \citep{ber08a}, auroral emission from the polar regions \citep{hal06}, and the presence of large coronal loops with widely separated magnetic footprints in a global dipolar magnetic field \citep{lyn15}.  In the auroral model, which was recently proposed to explain simultaneous radio flaring and H$\alpha$ activity on LSR J1835+3259 and other brown dwarfs \citep{hal15,kao16}, the rapid changes in radio flare polarity on TVLM 513 are caused by viewing angle effects, as magnetic activity is observed emanating from alternating polar caps in a global dipolar field that is tilted with respect to the object's rotational axis \citep{hal07}.  Both chromospheric and coronal models also rely on the rotation of non-polar magnetic spots or structures into and out of the field of view. The chromospheric and coronal regions above sunspots and flares on the Sun are known sites of radio emission with $\leq$3.5 kG strength magnetic fields present \citep{sol03}, thus similar in magnitude to those found on UCDs.

The observation of radio emission of a single magnetic orientation over months to years indicates that terrestrial observers have a restricted viewing angle of magnetic active regions. An observed change in the magnetic field polarity could indicate that magnetic activity near one pole of a global, dipolar magnetic field is more readily observed than near the other pole, and that this pole has reversed its magnetic polarity as occurs during an activity cycle.  This idea is supported observationally for fully convective stars by the simple, dipolar magnetic topology of the M4 dwarf V374 Peg, as revealed by Zeeman-Doppler Imaging (ZDI) \citep{mor08}.  Alternatively, a new active region of opposite magnetic orientation could have emerged and begun to flare by the ECMI mechanism, while the previous active region has ceased to exist, become quiescent, or continues to flare, but has migrated out of our line-of-sight. On the Sun, active regions exist on timescales of hours to months, with their areas decaying according to the Gnevyshev-Waldmeier rule \citep{sol03}.  That magnetic structures of a single observable polarity should endure for months to years, then entirely reverse polarity strongly suggests solar-like activity cycles on these substellar objects.

\section{Discussion}

These results should be viewed in the larger context of the main sequence on the Hertzsprung-Russell diagram.  Approximately 50\% of main sequence stars of types F2-M2 show evidence of magnetic cycles ranging in length from 2.5-25 years, with another $\sim$10\% exhibiting potential longer-term magnetic trends, as determined by the periodic changes in their chromospheric Ca II H and K emission \citep{bal95}.  More recently, photometric and Ca II observations of a handful of G0-M2 stars indicate that $\sim$75\% appear to host one or more activity cycles \citep{ola09}.  The next well-studied stellar object, after the Sun, with clear proof of a magnetic activity cycle, is the F7V $\tau$ Boo, as determined by two reversals of its ZDI-measured magnetic topology during its $\sim$2-yr activity cycle.  $\tau$ Boo is also noteworthy since it hosts an M$sin i = 4.1\pm0.153 M_{J}$ hot Jupiter with a 3.31-day orbital period at a distance of 0.049 AU, although searches for radio emission from any interaction have yielded no detection (Hallinan et al. 2013 and references therein).  These results indicate that magnetic reversals are pervasive among stars, including M dwarfs which may have planetary systems, with timescales not unlike those hypothesized for brown dwarfs.

Moving along the main sequence toward cooler temperatures, a decade-long study of chromospheric H$\alpha$ emission from K5-M5 stars found periodic H$\alpha$ variability in $\sim$6\% of the sample that ranged from 0.8-7.4 yrs in length, while an additional 8\% may have activity cycles longer than 10 years, i.e. the length of the survey work \citep{rob13}.  Such a study of H$\alpha$ variability may be extended to brown dwarfs.  For instance, in 2002 September, H$\alpha$ observations simultaneous with the radio observations of J0036+18 failed to detect emission down to $log(L_{H\alpha}/L_{bol})\leq$-6.65 \citep{ber05}, while clear H$\alpha$ emission at a level of $log(L_{H\alpha}/L_{bol})$=-6.1 was detected in 2012 July \citep{pin16}.  This H$\alpha$ variability may trace the progression of this L3.5 brown dwarf from solar minimum to solar maximum, a process with an approximately decadal timescale.

Based on ZDI measurements of 11 M5-M8 dwarfs, \citet{mor10} grouped these fully convective stars into two categories of magnetic field topologies: (1) strong, axisymmetric poloidal fields or (2) weaker, non-axisymmetric, poloidal fields with an additional significant toroidal component.  They speculated that this behavior could be explained by these low-mass stars switching betwixt two dynamo states.  \citet{kit14}, through analogy with solar behavior, provided the theoretical basis that these two types of magnetic topologies may actually be the result of a global axisymmetric field that oscillates on $\sim$100 year timescales. Their investigation of fully convective M dwarfs relied on an $\alpha^{2}\Omega$ dynamo model \citep{cha06} that had been modified to include differential rotation and decreased eddy magnetic diffusivity.  The examined evolution in UCD radio emission polarity and the change in H$\alpha$ behavior of J0036+18 suggest that this oscillatory magnetic field model may apply beyond M dwarfs at the end of the main sequence to brown dwarfs as well.

This work has argued for the presence of solar-like magnetic activity cycles in UCDs mainly on the basis of systematic changes in the circular polarization helicity of their radio emission.  It is important to note two underlying assumptions of this framework.  First, that UCDs have large-scale dipolar magnetic fields, as the Sun and M dwarfs do \citep{mor08,mci14}. Second, that changes in radio emission polarity are tied to alterations in the underlying magnetic field, and are not the result of plasma or magnetic field inhomogeneity propagation effects \citep{bog09}.  This latter assertion can be evaluated by comparing flaring radio emission helicities with known magnetic topologies of M dwarfs as revealed by ZDI (e.g. Morin et al. 2010).

The hypothesis that UCDs undergo magnetic activity cycles has important implications for the detection statistics of these objects.  Whereas UCDs undetected in the past have been thought to be either quiescent radio emitters, requiring precise unpolarized intensity (Stokes I) measurements for detection, or are simply ``not radio active,'' these results suggest a more complex picture.  UCDs may enter periods during which they do not exhibit flaring radio emission, but are only detectable by their quiescent emission, resembling the Sun at solar minimum.  On the other hand, UCDs undetected at radio wavelengths may indeed have emission that is too faint for detection or only infrequently accumulate enough plasma from the interstellar medium, leading to sporadic radio emission.  Although a substellar object may not be observed to emit radio waves during a particular observation epoch, the object may later be detected when it enters the solar maximum phase of its activity cycle.  Therefore, undetected sources should be periodically revisited to search for radio emission.

Whether UCD activity cycles exist can be determined by long-term monitoring of known radio emitters, which would include recording the radio flare helicity, the cutoff frequency of emission, and the overall radio flux from the source at various frequencies.  If a global dipolar field exists and reverses direction, this may be detectable by temporal changes in the cutoff frequency of the radio emission over month-to-year long timescales.  For example, the annual averages of sunspot magnetic field strengths vary in accordance with the solar cycle, with stronger fields at solar maximum that then decline until solar minimum, when magnetic reversal occurs \citep{pev14}.  A study of the monthly mean total stellar radio flux offers another means of assessing the presence of activity cycles in brown dwarfs, as this quantity varies in a prominent way during the solar activity cycle for 1-9.6 GHz frequencies \citep{shi11}. Similar observations of UCDs could be conducted, but would require broadband radio observations over a number of years.

The potential existence of magnetic activity cycles in objects with fully convective interiors poses serious challenges to theories of how magnetism is generated within stars.  The $\alpha\Omega$ dynamo, which is the most popular theory to explain magnetic activity on the Sun \citep{cha14}, requires the existence of a tachocline, the shearing interface between the radiative core and the convective envelope, to produce large-scale magnetic fields.  But as mid-late M dwarfs and brown dwarfs lack radiative cores, this theory would predict that large-scale magnetism should be absent, contradicting the observations presented.  Alternatively, $\alpha^{2}$ dynamos can produce large-scale, higher-order multipolar, kG-strength magnetic fields in fully convective objects, but the stability of the magnetic fields precludes shorter, decadal activity cycles (if they are permitted at all), as appears to exist for both M dwarfs and brown dwarfs \citep{cha06,rob13,kit14}.  The turbulent dynamo mechanism, which has also been proposed to explain magnetism in low-mass, fully convective objects postulates the existence of a low-magnitude magnetic field of mixed polarity on small spatial scales \citep{dur93}.  While this mechanism may power the quiescent radio and H$\alpha$ emission during activity cycle minima, it is insufficient to form active regions capable of producing kG-strength magnetic fields and large ECMI flares.  Recent modeling efforts that describe the formation and amplification of magnetic wreaths solely in the solar convective zone, which result in starspots and magnetic activity cycles \citep{nel14}, could, however, provide a promising means to explain the magnetic properties and activity cycles of fully convective low-mass stars and brown dwarfs.

\section{Acknowledgements}

This research has made use of NASA's Astrophysics Data System and the SIMBAD database, operated at CDS, Strasbourg, France.

\clearpage
\begin{deluxetable}{llllllllll}
\tabletypesize{\scriptsize}
\tablecolumns{10}
%\tablewidth{0pt}
\tablecaption{Radio Flaring UCD Polarity Observations}
\tablehead{
	\colhead{Object}&
	\colhead{Spectral}&
	\colhead{Period}&
	\colhead{Detection}&
	\colhead{Date}&
	\colhead{Polarization}&
	\colhead{Pol.}&
	\colhead{Cycle}\\
	\colhead{}&
	\colhead{Type}&
	\colhead{(hr)}&
	\colhead{(Period)Refs}&
	\colhead{(yyyy mmm dd)}&
	\colhead{(Orientation)\%}&
	\colhead{Ref}&
	\colhead{Length(yr)}
}
\startdata
J1314+13\tablenotemark{b} & M7+M7 & 3.89 & 1,2 (1) & 2009 Mar 25 & 24(L,R)\tablenotemark{a} & 1 & 6\\
 & & & & {\bf 2012 Mar 26} & {\bf 100(L)} & {\bf 2} & \\
 & & & & 2013 Apr 28 & 100(L) & 2 & \\
 & & & & 2013 May 04 & 57(L) & 3 & \\
\hline
DENIS 1048-3956 & M8 & - & 4 & 2002 May 16 & 100(R) & 4 & ?\\
\hline
LSR 1835+32 & M8.5 & 2.84 & 5 (5) & 2006 Sep 18 & 100(L) & 5 & 10\\
& & & & 2007 May 03 & 50(L)\tablenotemark{a} & 6 & \\
& & & & {\bf 2011 Jul 23 }& {\bf 100(R)} & {\bf 7} & \\
\hline
TVLM 513-46546 & M9 & 1.96 & 8 (9,10) & 2001 Sep 23 & 66(R) & 8 & 7\\
& & & & {\bf 2006 May 20} & {\bf 100(L,R)} & {\bf 10} & \\
& & & & 2008 Dec - 2013 May & 100(L,R) & 11 & \\
\hline
LP 944-20 & M9 & - & 12 & 2000 Jul 27 & 30(L,R) & 12 & ? \\
\hline
BRI 0021-0214 & M9.5 & - & 8 & 2001 May 31 & 30(R) & 8 & ? \\
\hline
J0746+20\tablenotemark{b} & L0.5+L1.5 & 2.07 & 13 (14) & 2007 Jan 22 & 100(L) & 13 & 3\\
& & & & 2008 Feb 22 & 100(L) & 14 & \\
& & & & {\bf 2010 Nov 12} & {\bf 50-100(R)} & {\bf 15} & \\
& & & & {\bf 2010 Dec 16} & {\bf 50-100(L),0-50(R)} & {\bf 15} & \\
& & & & 2012 Feb 08 & 73(L) & 11 & \\
& & & & 2012 Dec 14 & 50-100(L) & 15 & \\
\hline
J0036+18\tablenotemark{b} & L3.5 & 3.08 & 8 (5) & 2001 Oct 9 & 62(L) & 8 & 20\\
& & & & 2002 Sep 28 & 60-73(L) & 16 & \\
& & & & 2005 Jan 10 & 46(L) & 16 & \\
& & & & 2006 Sep 24 & 47(L) & 5 & \\
\hline
J0423-04\tablenotemark{b} & L6+T2 & - & 17 & 2013 May- 2013 Aug & 0-49(L),0-97(R) & 17 & ? \\
\hline
J1043+22\tablenotemark{b} & L8 & - & 17 & 2013 May 25-27 & 78(L) & 17 & ? \\
\hline
J0136+09\tablenotemark{b} & T2.5 & - & 17 & 2013 May 18 & 64(L) & 17 & ? \\
\hline
J1122+25\tablenotemark{b} & T6 & 0.288 & 18 (18) & 2013 May 8 & 15(L) & 18 & 1+ \\
& & & & 2013 Dec 27, 31 & 100 (L) & 18 & \\
\hline
J1237+65\tablenotemark{b} & T6.5 & - & 17 & 2013 May 21 & 28-49(R) & 17 & ?\\
\hline
J1047+21\tablenotemark{b} &  T6.5 & 1.77 & 19 (20) & 2010 Jan 8 & 89(R) & 19 & 7\\
& & & & 2010 Dec 5 & 72(R) & 19 & \\
& & & & 2011 Feb 7 & 18(R) & 19 & \\
& & & & 2013 May 6 & 33(R) & 3 & \\
& & & & {\bf 2013 May 19} & {\bf 75(L)} & {\bf 17} & \\
& & & & 2013 Dec 28 & 50-100(L) & 20 & \\
\enddata
\tablecomments{Properties and flaring history of periodically radio-flaring UCDs.  The helicity of the radio flares is indicated as either left (L) or right (R) circularly polarized.  Bold denotes changes in flare polarity that may indicate the presence of an activity cycle.  Note that many more observation epochs of TVLM 513 have been conducted from 2002 - 2013, see \citet{wol14} for details. Most magnetic activity cycle lengths are estimated from the available radio data; the activity cycle length for J0036+18 is crudely estimated from H$\alpha$ measurements. {\bf References.} (1) \citet{mcl11}; (2) \citet{wil15}; (3) \citet{rou16b}; (4) \citet{bur05}; (5) \citet{hal08}; (6) \citet{ber08b}; (7) \citet{hal15}; (8) \citet{ber02}; (9) \citet{hal06}; (10) \citet{hal07}; (11) \citet{rouphd}; (12) \citet{ber01}; (13) \citet{ant08}; (14) \citet{ber09}; (15) \citet{lyn15}; (16) \citet{ber05}; (17) \citet{kao16}; (18) \citet{rou16a}; (19) \citet{rou12}; (20) \citet{wilb15}.}
\tablenotetext{a}{No flares were observed during this time.}
\tablenotetext{b}{Full names: 2MASS J13142039+1320011 AB, 2MASS J07464256+2000321 AB, 2MASS J0036159+182110, SDSS 04234858-0414035, 2MASS 10430758+2225236, 2MASS 01365662+0933473, WISEPC J112254.73+255021.5, 2MASS 12373919+6526148, 2MASS 10475385+2124234, respectively.}
\end{deluxetable}

\end{document}